\documentclass[dvips]{article}

\usepackage{icrc2011}

\title{Orbit Mode observations of Crab and Mrk 421}

\newcommand{\etal}{\MakeLowercase{\textit{et al.}}} 
\shorttitle{Kieda \etal Orbit Mode Observations }

\authors{D.B. Kieda$^{1}$, for the VERITAS collaboration$^{2}$ }
\afiliations{$^1$Department of Physics and Astronomy, University of Utah, 115 S 1400 E, Salt Lake City, UT 84112 USA\\ $^2$see Holder et al (these proceedings)
}
\email{kieda@physics.utah.edu}

\abstract{ The canonical observation mode for IACT gamma-ray observations employs four discrete pointings in the cardinal directions (the "wobble" mode). For the VERITAS Observatory, the target source is offset by 0.5-0.7 degrees from the camera center, and the observation lasts 20 minutes. During January/February of 2011, the VERITAS Observatory tested a new "orbit" observation mode, where the target source is continuously rotated around the camera center at a fixed radial offset and constant angular velocity. This mode of observation may help better estimate the cosmic ray background across the field of view, and will also reduce detector dead-time between the discrete 20 minute runs. In winter 2011, orbit mode observations where taken on the Crab Nebula and Mrk 421. In this paper we present the analysis of these observations, and describe the potential applications of orbit mode observations for diffuse (extended) sources as well as GRBs.  }

\begin{document}
\maketitle

\section{The VERITAS Imaging Atmospheric Cherenkov Telescopes}
VERITAS \cite{holder}\cite{perkins}, located at the Fred Lawrence Whipple Observatory (FLWO) in southern Arizona, USA, is an array of four 12 meter diameter Imaging Atmospheric Cherenkov Telescopes (IACT).   VERITAS can detect gamma-rays with energies from 100 GeV to 30 TeV with a flux of one percent of the Crab Nebula in approximately 25 hours. VERITAS has an energy resolution of 15-25\%, an angular resolution of 0.1 degrees (68\% containment radius), and a pointing accuracy within 50 arc-seconds.

\section{Source Locations Reconstruction}
\begin{figure}[!t]
  \vspace{5mm}
  \includegraphics[width=3in]{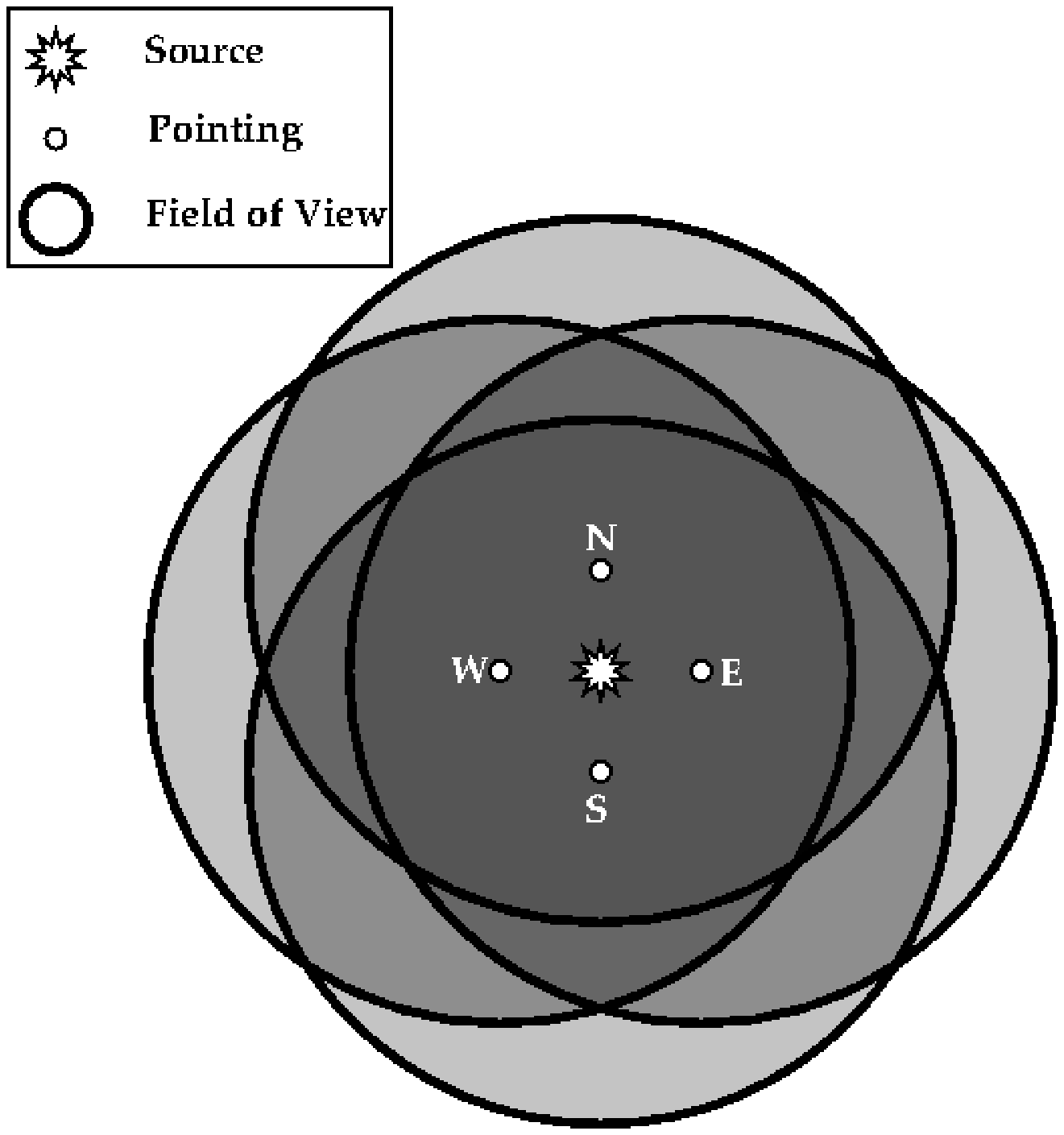}
  \includegraphics[width=3in]{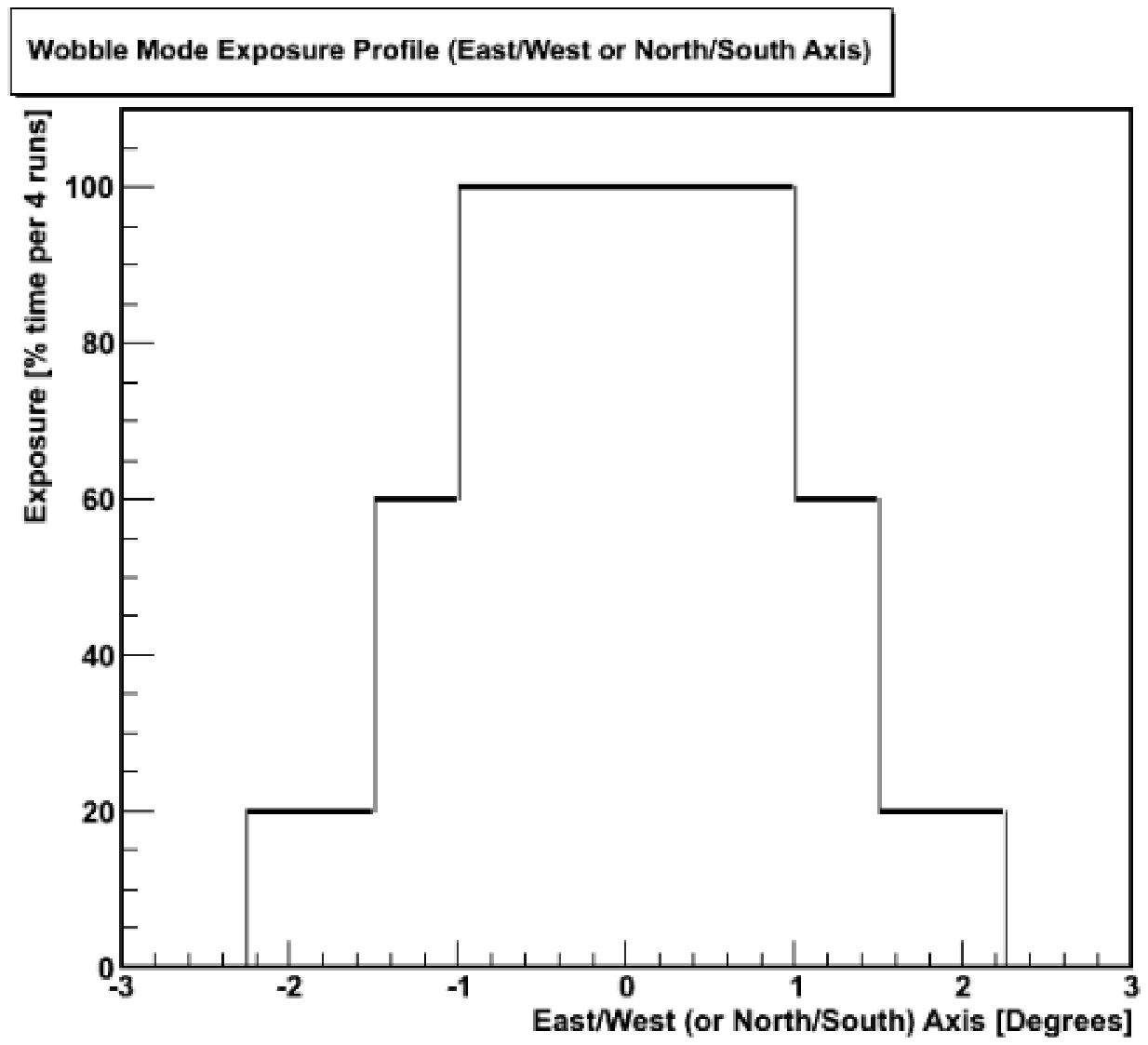}
  \caption{Exposure area (top) of the wobble mode technique for four runs with East to West (or North to South) profile (bottom)  }
  \label{wobble_fig}
 \end{figure}

\begin{figure}[!t]
  \vspace{5mm}
  \hspace{-5mm}
  \includegraphics[width=3.0in]{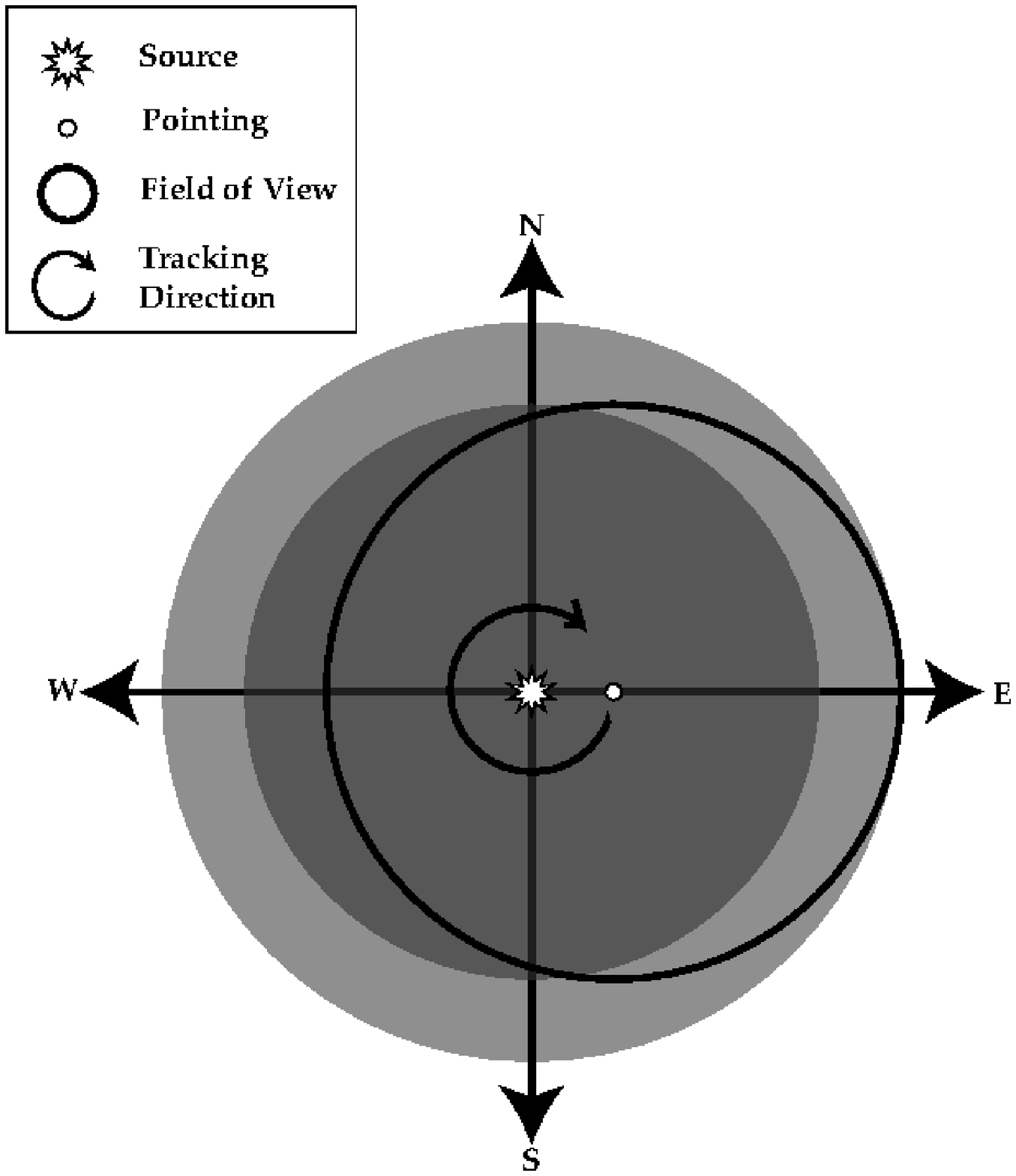}
  \includegraphics[width=3.0in]{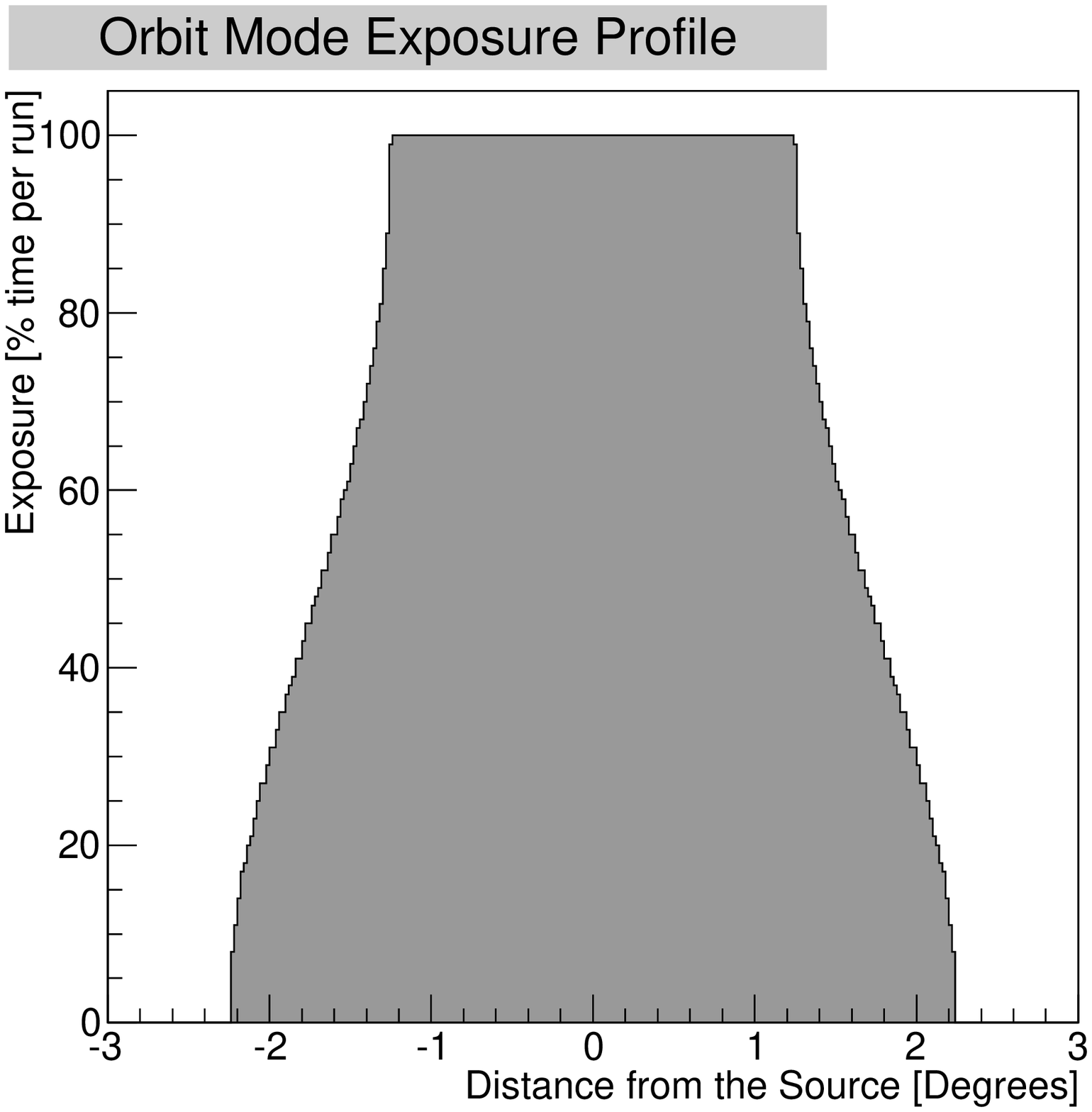}
  \caption{Exposure area (top) of the orbit mode technique with  profile (bottom)   }
  \label{orbit_fig}
\end{figure}

\begin{figure}[!t]
  \vspace{5mm}
  \includegraphics[width=2.75in]{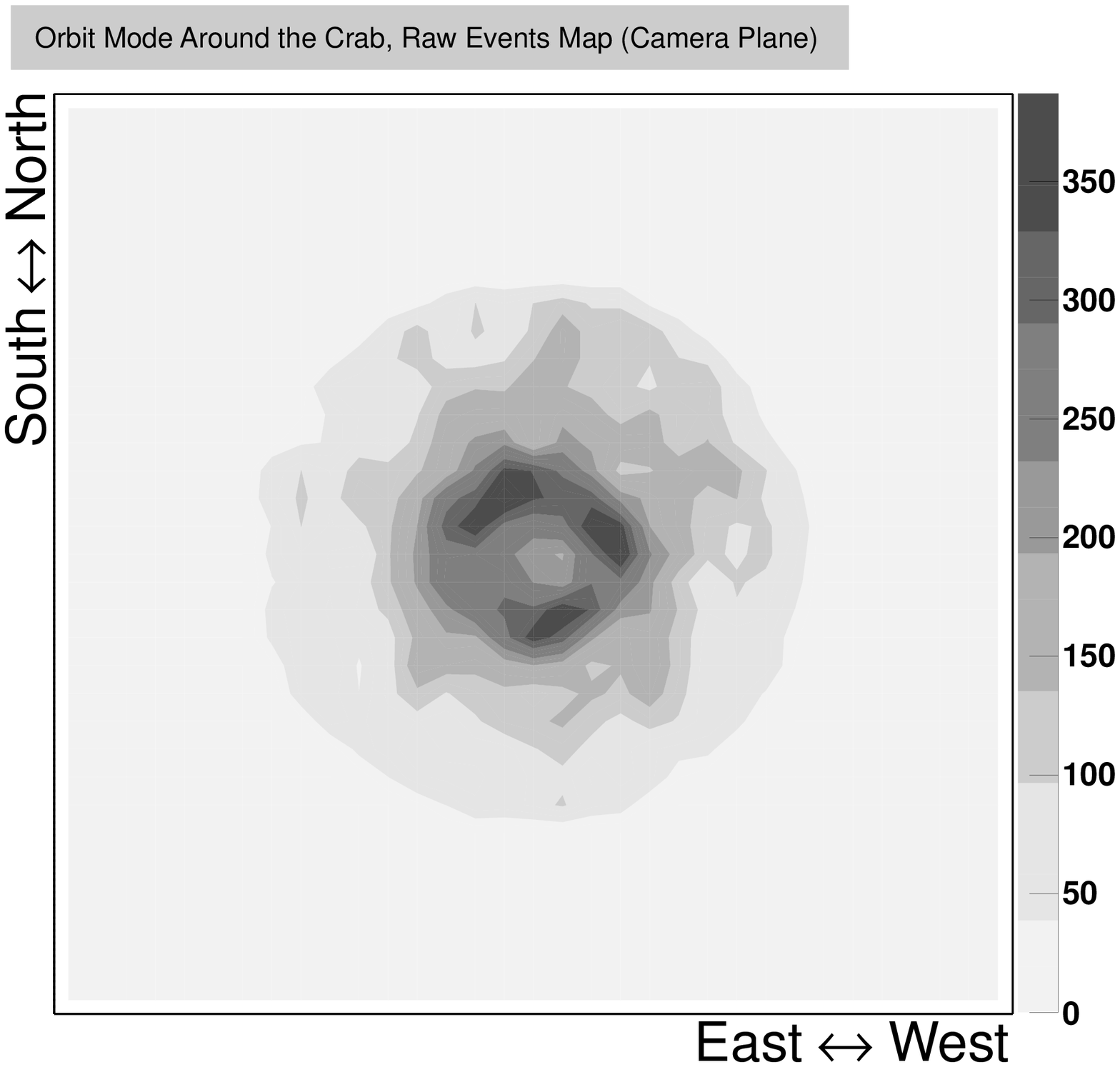}
  \caption{Image of the Crab Nebula before rotation corrections are applied and before the background subtraction}
  \label{crab_raw}
\end{figure}

\begin{figure}
  \vspace{25mm}
\includegraphics[width=3in]{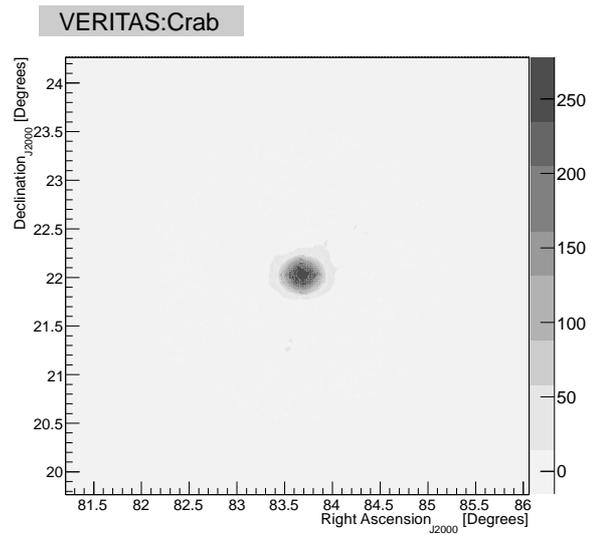}
  \caption{Excess counts for one thirty minute run of the Crab Nebula using the orbit mode observation technique}
  \label{crab}
\end{figure}

\begin{figure}[!t]
  \vspace{5mm}
  \includegraphics[width=3in]{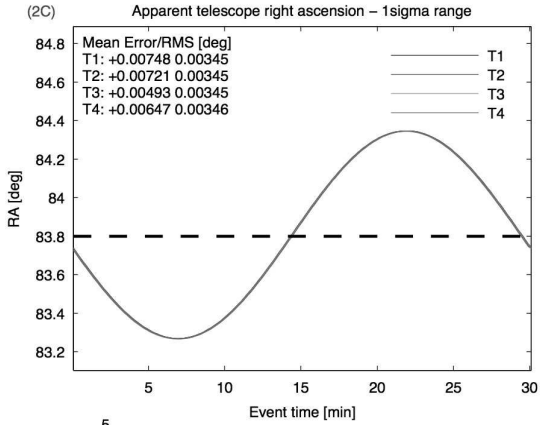}
  \caption{Apparent right ascension of the telescopes during an orbit mode observation}
  \label{sine}
\end{figure}

\begin{figure}[!t]
  \vspace{5mm}
  \includegraphics[width=3in]{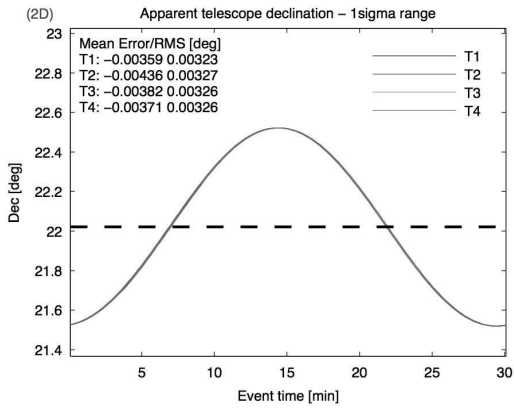}
  \caption{Apparent declination of the telescopes during an orbit mode observation}
  \label{cosine}
\end{figure}

\begin{figure}[!t]
  \vspace{5mm}
  \includegraphics[width=1.75in]{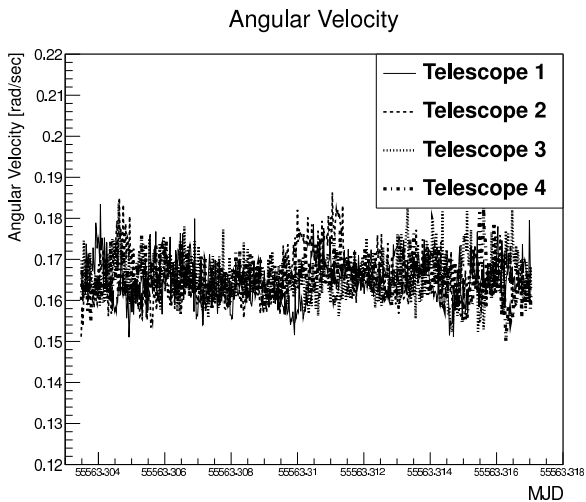}
  \caption{Angular velocity of the telescopes during an orbit mode observation}
  \label{angular_vel}
\end{figure}

\begin{figure}[!t]
  \vspace{5mm}
  \includegraphics[width=1.75in]{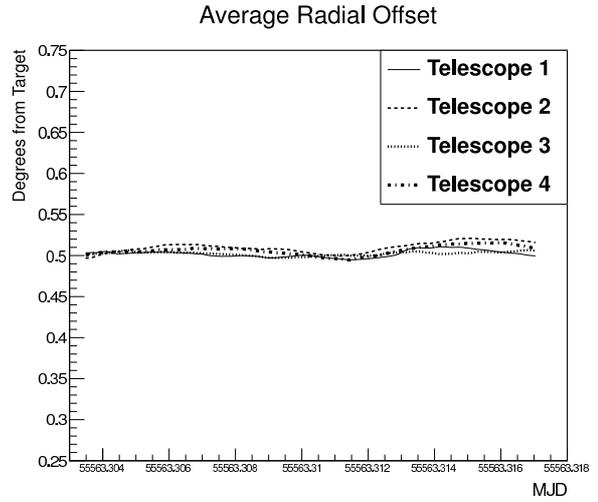}
  \caption{Radial offset of the telescopes during an orbit mode observation}
  \label{radial_off}
\end{figure}

VERITAS observations are normally taken in wobble mode\cite{Daum}. During an observation using wobble mode, the center of the camera is held at a fixed position in right ascension and declination offset from the intended targeted source [See Figure \ref{wobble_fig}]. In orbit mode the center of the camera circumscribes the source in right ascension and declination with an angular velocity and radial offset dependent on the type of source (point-like, extended, or a GRB). Typical values for a point-like source are one revolution per 20 to 80 minutes and a 0.5 degree radial offset [See Figure \ref{orbit_fig}]. Using orbit mode,  prior to rotation corrections to the field of view, the source appears as a ring [See Figure \ref{crab_raw}]. For each event there is an elevation and azimuth angle recorded. With this information and the elevation and azimuth of the telescopes, the reconstructed direction can be found [See Figure \ref{crab}]. Figures \ref{sine} and \ref{cosine} show the pointed position (in right ascension and declination as a function of time) of each telescope followed a smooth sine and cosine curve. During testing it has been shown that the angular velocity and radial offset are fairly constant [See Figures \ref{angular_vel} and \ref{radial_off}]. 

\section{Discussion}
\begin{figure}
 \vspace{5mm}
  \includegraphics[width=2.9in,height=3in]{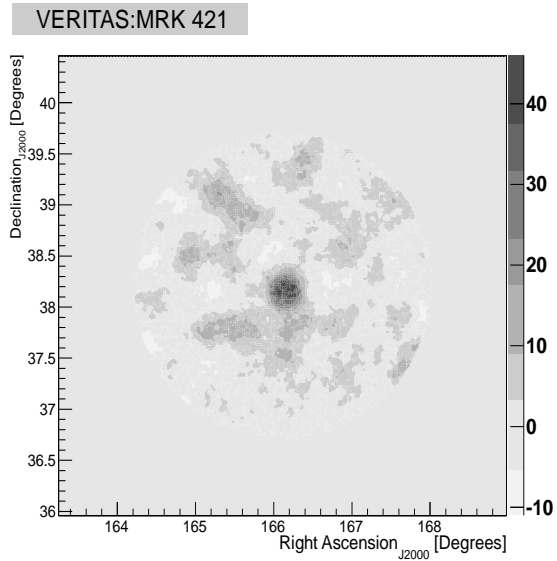}
 
 \caption{Excess counts for one eighty minute run of Mrk 421 using the orbit mode observation technique}
  \label{mrk421}
\end{figure}

The orbit mode technique was developed to help eliminate dead-time during transitions between wobble directions for data runs sets, to slightly increase the area of the field of view by maintaining azimuthal symmetry of the exposure around the source, and to produce an uniform background estimate. In order to minimize the dead-time between runs, we had to test whether the VERITAS data network could transfer file sizes of twenty to thirty gigabytes. This was successfully done with a run during the daytime with the charge injection system of the telescopes, and later on single eighty minute data run of Mrk 421 [See Figure \ref{mrk421}]. The typical time between data runs for slewing of the telescopes can last one to two minutes. If implemented for regular data operations, orbit mode would add additional thirty to sixty minutes of observation time per night. 

Orbit mode has been developed to test whether the background estimation using the reflected regions method\cite{Aharonian} or ring background method\cite{Berg} would be more uniform and therefore increasing the sensitivity of the analysis. Preliminary results of the orbit mode analysis on the Crab Nebula produced $10.0 \pm 0.6$ gamma-rays a minute. A wobble mode analysis was also performed on the Crab Nebula with data taken the same night at a similar zenith angle produced $9.1\pm 0.7$ gamma-rays per minute. 

 Stars in the field of view can cause higher trigger rates in individual pixels, and therefore cause a higher background level in small regions of the sky. With the stars more rapidly rotating in the field of view during orbit mode observations than during wobble mode observations, this may eliminate some of the background noise in the small regions of the sky effected by the stars. 

GRB alerts have a large error associated with the position. A GRB alert from the Fermi\cite{fermi} LAT or Fermi GBM\cite{fermi} has an error of approximately $2^{\circ}$ or $15^\circ$ (radius, one sigma) respectively. Orbit mode allows a rapid scan of a larger area (with a larger radial offset) of the sky than wobble mode. For a Fermi LAT alert, the entire one sigma containment radius can be observed in orbit mode with VERITAS in one orbit. For a Fermi GBM alert, approximately twenty-five percent of the one sigma containment radius can be observed with VERITAS in one orbit using orbit mode.

\section{Conclusion}
Orbit mode observations have been successfully tested on the VERITAS telescopes. The telescopes have performed exceptionally well with the orbit observation mode. Preliminary results of the Crab Nebula have already qualitatively demonstrated that orbit mode observations have a similar or better gamma-ray detection rate to wobble mode observations for the Crab Nebula, a point source. Further observations and simulations are underway to quantify the comparison between the two observation modes, as well as determine the sensitivity of orbit mode for extended, diffuse gamma-ray sources.

\section{Acknowledgments}
This research is supported by grants from the US Department of Energy, the US National Science Foundation, and the Smithsonian Institution, by NSERC in Canada, by Science Foundation Ireland, and by STFC in the UK. We acknowledge the excellent work of the technical support staff at the FLWO and the collaborating institutions in the construction and operation of the instrument.

\clearpage

\end{document}